\documentclass[amssymb,11pt,epsf]{article}
\usepackage{graphicx}
\usepackage{amssymb}

\usepackage{graphicx}

\begin{document}

\author{Alexandre Dom\'{\i}nguez-Clarimon \footnote{landi@ecm.ub.es} \\
Departament d'Estructura i Constituents de la Mat\`eria,\\
Facultat de F\'{\i}sica, Universitat de Barcelona\\
Diagonal, 647, 08028 Barcelona, Spain}

\title{Decoherence of a particle across a medium:\\
 a microscopic derivation}
\date{}

\maketitle

\begin{abstract}

We deduce from a microscopic point of view the equation that describes how the state of a particle crossing a medium decoheres. We apply our results to the example of  a particle crossing a gas, computing explicitly the Lindblad operators in terms of the interaction potential between the particle and a target of the medium. We interpret the imaginary part of the refraction index as a loss of quantum coherence, that is reflected in the disappearance of  interference patterns in a Young experiment.
\end{abstract}

\vfill
\vbox{
UB-ECM-PF 03/08\null\par
March 2003}

\section{Introduction}

When a particle crosses a medium the loss of quantum coherence in its state is unavoidable. This means that the statistical mixture of the particle state increases as it moves on. There are two ways of increasing such mixture. One of them is purely classical and comes from a  probabilistic knowledge of which the state of the medium is. The other one is specific of Quantum Mechanics (QM) and it requires a quantum description for the medium too. Here, decoherence appears because the particle gets entangled with the medium through the interaction with it.

Some of these effects that the medium exerts on the particle have long been understood and well described by means of an effective hamiltonian, often expressed as an index of refraction of the medium for the particle propagation. This refraction index was introduced by Fermi \cite{Fermi} in the context of neutron dispersion by matter, and is of the form:
\begin{eqnarray}
\frac{k'}{k}=1+2\pi n \frac{f(k,k)}{k^2} \label{Fermiequation}
\end{eqnarray}
where $f(k,k)$ is the forward scattering amplitude of the particle by a single dispersor center and $n$ is the density of targets. It closely follows the calculation of the refraction index for light in Classical Electrodynamics when it propagates through a dielectric medium, first derived by H.A.Lorentz  (see \cite{Jackson}).

These corrections induced by the medium to the free evolution are crucial in the effects that matter has on neutrino oscillations \cite{MSW, Wolfenstein}  (relevant for the  MSW effect), and on kaons \cite{Bramon, kabir}.

One feature of the refraction index type of correction is that the evolution of the particle state is still hamiltonian and, thus, its coherence is strictly preserved in time. On the contrary, this article is focused on the decoherence effects that are not included in a refraction index.

A detailed study of the decoherence process would, of course, require a tracking of the whole system, including the environment besides the particle, which would follow a standard QM evolution. It is when one disregards the environment and focuses just on the particle state that its dynamics departs from the one described by the usual Schr\"odinger equation. It becomes non-local in time, i.e., it depends on the entire history
of its evolution. Under certain conditions, this non-locality is very weak and decoherence can be effectively treated  by adding new, non-hamiltonian, terms to the differential equation for the density matrix. (Notice that here, as decoherence is at work, the state of the system  is naturally described in terms of a density matrix rather than a wave function). Such differentiall equation is known as the Lindblad equation and is of the form:
\begin{eqnarray}
\dot{\rho}=-i[H_{eff},\rho ] + \sum_j \left( 2A_j \rho A_j^{\dagger} -  A_j^{\dagger}A_j \rho -\rho A_j^{\dagger}A_j \right)\label{Lindblad proved}
\end{eqnarray}
The presence of an environment has two effects: it modifies in part the hamiltonian and it moreover generates the operators $A_i$ which make quantum coherence of the particle itself decrease irreversibly. Lindblad proved equation (\ref{Lindblad proved}) with the assumption that the evolution of the density matrix in a decoherent such process fulfills \textit{quantum dynamical, completely positive, semigroup} sort of composition law, which increases entropy \cite{Lindblad}. This is a  very general treatment and has been widely used to describe many different physical situations. Examples of these are  Quantum Optics \cite{Louisell, Cohen-Tannoudji} (where it appeared for the first time);  the time dependence of the optical activity in chiral isomer molecules \cite{Stodolsky} ;  the emergence of a classical description for the macroscopic objects as well as the attempts to explain the mechanism of some issues of foundation in QM \cite{Zurek, Zeh}. 

%

Recently, it has also been the framework to address the question if decoherence could blur the oscillations in the experiments of neutral kaons \cite{peskin, benatti, taron}, or the evidences of neutrino oscillations (\cite{review} and references therein).

In this article we explore the loss of coherence for a particle as it crosses a diluted medium.  We assume the interaction of the particle with each target is short range.  For this reason memory effects are washed out on time-scales larger than the interaction time. We use the scattering $S$ matrix to describe the interaction, since our coarsed grained time is infinite with respect to the intercation scale, for all purposes. A Lindblad equation thus naturally emerges.

We discern two different sources of decoherence: \textit{foootprint} and \textit{mixture}, each one contributing to distinct Lindblad operators.  We give explicit expressions for these operators in terms of the particle-target  potential, their masses and the target wave functions. In particular we  compute the  corrections to Fermi's equation (\ref{Fermiequation}) in terms particle and target mass ratio. We also see how the unitarity of this equation is restored once all pieces of decoherence in the Lindblad equation are properly included. Finally, we argue that the presence of a medium washes out the  interference pattern of a double slit Young experiment.

\section{The decoherence mechanisms: the footprint and the mixture }

We introduce, in this section, the basic ideas necessary to understand which
are the microscopic mechanisms that produce decoherence in the
case of a particle crossing a medium. We classify the sources of
decoherence in two types: the \textit{footprint} and the \textit{mixture}. Let us
clarify this distinction with the help of a toy model that consists of a particle and a box. The particle will pass through the box and will interact with it. For simplicity, suppose
they both have two-dimensional Hilbert spaces. The orthonormal
basis for the particle is $\{ |1\rangle , |2\rangle \}$, and for
the box is $\{ |a\rangle , |b\rangle \}$. After the particle
crosses the box, we do a partial trace over the box degrees of
freedom.
The two distinct mechanisms that we envisage are the following: \label{toy model}\\

\noindent \textbf{\textit{The footprint}}. Suppose that the
initial state is  $|1\rangle|a\rangle$. Suppose also
that after they have interacted the final state is $\frac{1}{\sqrt{2}}
\left(|1\rangle |a\rangle -|2\rangle |b\rangle \right)$, which is
entangled. The corresponding reduced density matrix for the
particle is the identity, which is not a pure state anymore. After
the interaction occurs, the box ''knows'' what the out state for
the particle is. The particle has left a footprint: if
the box is in state $|a\rangle$, the particle is in $|1\rangle$
and if the box is in $|b\rangle$ , the particle is in $|2\rangle$.
With the partial trace we overlook this information,
that remains in the box as a footprint. \\

\noindent \textbf{\textit{The mixture}}. Now we start with the
particle in $|1\rangle$, but the box in the mixed state
 $\frac{1}{2}|a\rangle \langle a|+\frac{1}{2}|b\rangle\langle b|$.
Choose an  interaction  as follows:
\begin{eqnarray}
&&|1\rangle \left\vert a\right\rangle \rightarrow |1\rangle
\left\vert a\right\rangle ~~~~~~|2\rangle \left\vert
a\right\rangle \rightarrow |2\rangle \left\vert a\right\rangle\\
&&|1\rangle \left\vert b\right\rangle \rightarrow |2\rangle
 \left\vert b\right\rangle ~~~~~~~|2\rangle \left\vert b\right\rangle \rightarrow |1\rangle
\left\vert b\right\rangle.
\end{eqnarray}
Notice that with this interaction and this initial state, the box
does not have any ''knowledge'' about what the out state for the
particle is: regardless of whether the box starts in $|a\rangle$
or $|b\rangle$,  it remains unchanged. Yet, decoherence also
appears: the reduced density matrix for the particle is the
identity again. The source for decoherence is, in this case, the initial mixed state for the box .\\

\section{A model for the medium} \label{aproximacions sobre el medi}

In this section we put forward the approximations that we make, as
well as the procedures that will eventually lead to a differential
equation for the particle reduced density matrix as it crosses the
medium.

Our medium is made of just one kind of targets. There is neither
interaction between them nor overlapping of their wave functions.
We also consider that the interaction between the particle and the
medium is weak; therefore, we neglect terms higher than second
order in the potential.

Our procedure consists in dividing the medium in thin slabs of
matter that will be crossed one by one by the particle. Each time
the particle crosses a slab, we do a partial trace over the slab
degrees of freedom and obtain a step by step equation for the
particle reduced density matrix. Eventually, we will get a differential
equation in time for it.

The targets of each slab are in a mixed state: target $j$ is in
state $|m_{j}\rangle$ with probability $q_{m_{j}}$. Since the
medium is homogeneous, the states of the different targets are
related simply by translations, and the corresponding weights are
the same.

If the initial state is pure $|in\rangle = |particle\rangle
|slab\rangle$, after the particle crosses the slab the system is
in state
\begin{eqnarray}
|out\rangle \simeq (1+i \sum_{j=1}^{N}T^{~\left( j
\right)})|particle\rangle |slab\rangle \label{aproximacio de la
matriu de scatering}
\end{eqnarray}
were $T^{~\left( j\right) }$ is the scattering operator of the
particle with target $j$ only, and $N$ is the number of targets in the slab.

 In order to justify this approximation, let us look at the case
where there is just one particle and two targets, and consider the
amplitude of the process of a particle that goes from $x$ to $y$,
and targets go from $x_{a},x_{b}$ to $y_{a},y_{b}$, in a time
interval $t$. The terms to second order in the potential are those
shown in \textit{Fig 1}. We must integrate over the coordinates
and the instants in which the interactions take place. Recall that
the free propagator is proportional to the phase factor
$\exp(im\frac{(y-x)^{2}}{2t})$, $m$ being the mass of the particle.
When we integrate over the internal coordinates, $z$ for instance,
the oscillation of the exponential is much faster in picture $1.b$
than in picture $1.a$ . This is so because the size of the region
where a particle and a target do interact is much smaller than the
distance among targets. Therefore, we can neglect those terms of
$1.b$ in front of those in $1.a$ . Ultimately, this is the same as
(\ref{aproximacio de la matriu de scatering}).

\begin{figure}[!h]
\includegraphics[scale=0.55,angle=0]{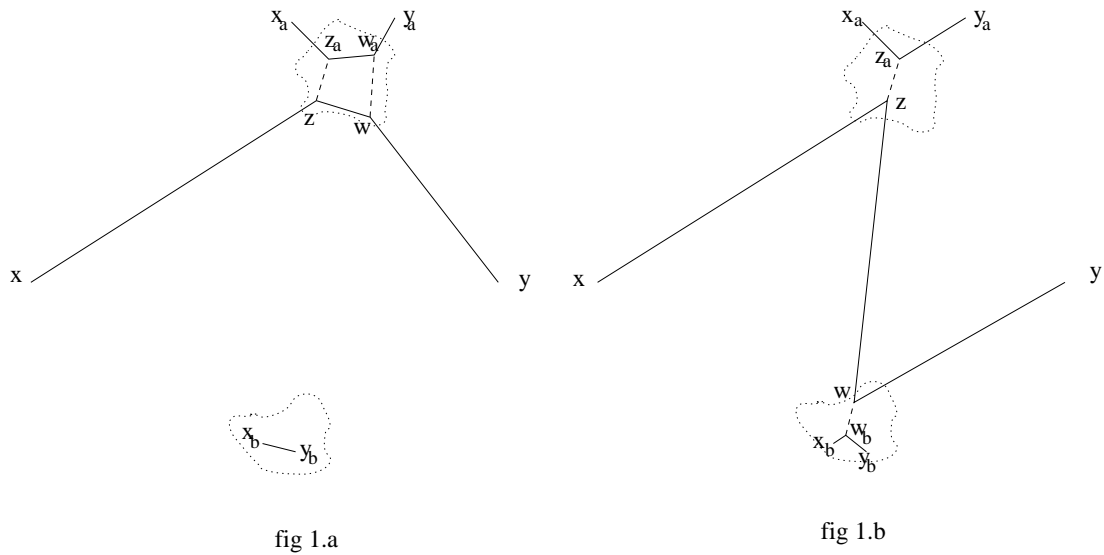}
\caption{The two order terms in $V$ for the $T$ matrix}
\end{figure}

It is worth pointing out that unitary relations hold both
for each $1+i T^{\left( j\right)}$ and for \break
$1+i\sum_{j=1}^{N}T^{~\left( j \right)}$ (within the
approximations above explained). Thus, terms like the crossed
product $T^{\dag \left( i\right)}T^{\left( j\right)}$ are
neglected by consistency.

\section{The one step equation}

\label{seccio one step}

In this section we obtain the particle density matrix after it has
crossed one slab. In the next section we will iterate this one
step evolution $r$ times.

If the initial state for the particle is $|\phi^{n}\rangle$ and for the slab is $|m\rangle \equiv
\prod_{j=1}^{N} |m_{j}\rangle$, the out-coming state for both together is: 
\begin{eqnarray}
|\textrm{out}^{n,m}\rangle &\simeq&(1+i \sum_{j=1}^{N}T^{~\left(
j\right) })|\phi^{n}\rangle |m\rangle, \label{out state}
\end{eqnarray}
The reduced density matrix for the particle is obtained by performing a partial trace over the slab:
\begin{eqnarray}
\rho _{red}^{n,m} =\textrm{TR}_{slab}\left[ |
\textrm{out}^{n,m}\rangle \langle \textrm{out}^{n,m}| \right]
\label{tracaparcial a lo bruto}
\end{eqnarray}
In general, the initial state will be mixed, and will have $|\phi^{n}\rangle$
with probability $p_n$ and $\prod_{j=1}^{N} |m_{j}\rangle$ with join
probability $\prod_{j=1}^{N} q_{m_j}$. Then, an average over all
possible initial states is due in (\ref{tracaparcial a lo
bruto}).

\subsection{The footprint in the slab and the reduced density matrix for the particle}
\label{the fotprint in the slab}

Let us  separate from  $|\textrm{out}^{n,m} \rangle $ the part
that contains the footprint left by the particle. This part can be
understood as the change that the particle leaves in the slab. It
is thus natural to define it with the help of the following
projector:
\begin{eqnarray}
\mathrm{P}= \mathrm{I} \otimes |m\rangle \langle m|\label{projector}
\end{eqnarray}
In order to separate the footprint term, we apply the projector
(1-P) to $|\textrm{out}^{n,m}\rangle$, which only keeps the part
of the slab that has changed:
\begin{eqnarray}
(1-\mathrm{P})|\textrm{out}^{n,m}\rangle \equiv
|\mathrm{footprint}^{n,m}\rangle \label{footprint state}
\end{eqnarray}
The part of $|out^{n,m}\rangle$ that leaves $|m\rangle$ intact is:
\begin{eqnarray}
\mathrm{P}|\textrm{out}^{n,m}\rangle  \equiv |\phi ^{^{\prime
}n,m}\rangle |m\rangle \label{phi prima}
\end{eqnarray}
One can thus write (\ref{out state}) as a sum of two orthogonal
terms:
\begin{eqnarray}
|\textrm{out}^{n,m}\rangle =|\phi ^{^{\prime }n,m}\rangle
|m\rangle
 + |\mathrm{footprint}^{n,m}\rangle,
\label{separem l'estat out}
\end{eqnarray}
and with it  (\ref{tracaparcial a lo bruto}) reads:
\begin{eqnarray}
\rho _{red}^{n,m} = | \phi ^{^{\prime }n,m}\rangle \langle \phi
^{^{\prime }n,m}| +\textrm{TR}\left[  |\mathrm{footprint}^{n,m}\rangle
\langle \mathrm{footprint}^{n,m} | \right]\label{separat}
\end{eqnarray}
In this way we separate the part in which the particle changes and
the slab does not (the first term), from the one which contains
any change produced in the slab (the second term). It is worth
pointing out that the second term in (\ref{separat}) encodes the
entanglement between the slab and the particle.
From (\ref{out state}) and (\ref{phi prima}):
\begin{eqnarray}
| \phi ^{^{\prime}n,m}\rangle &=&| \phi ^{n}\rangle +i
\sum\limits_{j=1}^{N}\langle m_{j}|T^{\left( j\right)
}|m_{j}\rangle | \phi ^{n}\rangle.
\end{eqnarray}
In terms of the operators $D_{M}^{\left( m_{j} \right)}\equiv
\langle m_{j}|~T^{~\left( j\right) }~|m_{j}\rangle$ (which act
only on the particle Hilbert space) the first term in (\ref{separat}):
\begin{eqnarray}
| \phi ^{^{\prime }n,m}\rangle \langle \phi ^{^{\prime
}n,m}|=\left( 1+i\sum\limits_{j=1}^{N}D_{M}^{\left( m_{j} \right)
}\right) | \phi ^{n}\rangle \langle \phi ^{n}| \left(
1-i\sum\limits_{j=1}^{N}D_{M}^{\dag \left( m_{j} \right) }\right)
\end{eqnarray}
As for the footprint term, from  (\ref{out state}) and
(\ref{footprint state}):
\begin{eqnarray}
|\mathrm{footprint}^{n,m}\rangle &=& i \sum\limits_{j=1}^{N} \left(
T^{~\left( j \right) }-\langle m_{j}|~ T^{~\left( j\right) }~|
m_{j}\rangle \otimes I^{(\otimes N)}\right)| \phi ^{n}\rangle
\prod\limits_{i=1}^{N} |m_{i}\rangle%
\end{eqnarray}
Each term ''$j$'' of this sum is orthogonal to any state which has target
$j$ in state $ | m_{j}\rangle $. For this reason, in the second
term of (\ref{separat})
\footnotesize
\begin{eqnarray}
\sum_{j,k}\textrm{TR}\left[ \left( T^{\left( j\right)}-\langle
m_{j}| T^{\left( j\right) }| m_{j}\rangle \otimes I^{(\otimes N)}
\right)|\phi^{n}\rangle \prod\limits_{i=1}^{N}| m_{i}\rangle
\langle \phi ^{n}| \prod\limits_{l=1}^{N}\langle m_{l}|\left(
T^{\dag \left( k\right)}-\langle m_{k}| T^{\dag \left( k\right) }|
m_{k}\rangle \otimes I^{(\otimes N)}\right) \right] \label{traca
parcialissima}
\end{eqnarray}
\normalsize
only the $j=k$ terms do contribute. In order to evaluate the
partial trace we introduce as a basis for the
Hilbert space of the $j$ particle  \{$|l_j\rangle$\}; (\ref{traca parcialissima})
then reads:
\footnotesize
\begin{eqnarray}
\sum_{j} \sum_{l_j} \left( \langle l_j | T^{\left( j\right)}|
m_{j}\rangle-\langle m_{j}| T^{\left( j\right) }| m_{j}\rangle
\langle l_j | m_{j}\rangle \right) |\phi^{n}\rangle  \langle \phi
^{n}| \left(\langle m_{j}| T^{\dag \left(j
\right)}|l_j\rangle-\langle m_{j}| T^{\dag \left( j\right) }|
m_{j}\rangle \langle m_{j}|l_j\rangle \right) 
\end{eqnarray}
\normalsize
which in terms of the operators $A_{E}^{(l_j,m_{j})}\equiv \langle
l_j| ~T^{~\left( j\right) }~| m_{j}\rangle - \langle m_{j}|
T^{\left( j\right) }| m_{j}\rangle \langle l_j| m_{j}\rangle$ is:
\begin{eqnarray}
\sum_{j,l_j}A_{E}^{(l_j,m_{j})}| \phi ^{n}\rangle \langle \phi
^{n}| A_{E}^{\dag (l_j,m_{j})}
\end{eqnarray}
Finally, we average in (\ref{separat}) over all possible initial states, for both the
particle and the slab, with their corresponding
probabilities\footnote{The unitary relations of the next section
guarantee the correct normalization of $\rho _{red}^{out}$}.
\begin{eqnarray}
\rho _{red}^{out}=&&\sum_{\{m_{j}\}}\prod_{j}q_{m_{j}}%
\left(1+i \sum_{i}D_{M}^{\left( m_{i} \right) }\right)
\rho_{red}^{in}%
\left(1-i\sum_{i}D_{M}^{\dag \left( m_{i} \right) }\right) \nonumber \\
+&&\sum_{\{m_{j}\}}\prod_{j}q_{m_{j}}%
\left(\sum_{i}A_{E}^{(l_i,m_{i})}\rho _{red}^{in} A_{E}^{\dag
(l_i,m_{i})}\right)\label{one-step-eq}
\end{eqnarray}
where $\rho_{red}^{in}=\sum\limits_{n} p_n | \phi ^{n}\rangle
\langle \phi ^{n}| $. We call this \textit{the one step equation}.
Notice that the footprint operators $A_{E}^{(l_i,m_{i})}$ are
independent of  the probabilities $q_{m_j}$.

For the slab we obtain an equation similar to (\ref{one-step-eq}):
\begin{eqnarray}
\rho _{slab}^{out}= \sum_{n}p_{n}%
\left(1+i \textrm{D}_{M}^{n} \right)%
\rho_{slab}^{in}%
\left(1-i \textrm{D}_{M}^{\dag n} \right)\nonumber \\
+ \sum_{n}p_{n}%
\left(\sum_{l}\textrm{A}_{E}^{(l,n)}%
\rho _{red}^{in} \textrm{A}_{E}^{\dag (l,n)}\right), \label{dens
matrix for medium}
\end{eqnarray}
It is interesting to note that if $\rho _{slab}^{in}$ is the identity, and
$[\textrm{A}_{E}^{\dag (l,n)},\textrm{A}_{E}^{(l,n)}]=0$, it 
remains unchanged\footnote{There is similar relations between
$\textrm{D}_{M}^{n}$ and $\textrm{A}_{E}$ that comes from unitary
relations like in (\ref{unitarirelations}).}. This may seem
contradictory, because footprint effects still appear in
(\ref{one-step-eq}), given that $A_{E}^{(l_i,m_{i})}\neq 0 $, whereas the medium has not changed. This is fixed by realizing that the initial state for the slab is a
mixture of pure states, and the footprint is left in each one of
them. By averaging over these pure states one erases the footprint
that the particle leaves on the slab. Yet, the footprint effects
on the particle are still there.

\subsection{Unitarity relations} \label{unitary relations}

Before we finish this section, let us put forward some unitarity
relations among the operators  $A$'s and $D$'s defined above. In
order to simplify the notation we omit the subindex $j$, and we
write $D_{M}^{\left( m \right) }$ and $q_{m}$ instead of
$\sum_{i}D_{M}^{\left( m_{i} \right) }$ and $\prod_{j}q_{m_{j}}$.
Unitarity of $S$ matrix reads:
\begin{eqnarray}
T-T^{\dagger }= i T^{\dagger }T \label{optical theorem}
\end{eqnarray}
Let us define the operator $D_E^{(m)}$ such that:
$T=D_{M}^{(m)}\otimes I^{(\otimes N)}+D_{E}^{(m)}$. Notice that then
$\langle l_j | D_E |m_j \rangle$ is nothing but $A^{l_j,m_j}_E$. We can rewrite
(\ref{optical theorem}) with the help of this decomposition:
\begin{eqnarray}
\left(%
D_{M}^{(m)} - D_{M}^{\dagger \,\,(m)}
- i D_{M}^{\dagger\,\,(m)}D_{M}^{(m)}\right) \otimes I^{(\otimes N)}%
+\left(  D_{E}^{(m)} -  D_{E}^{\dagger
\,\,(m)} - i D_{E}^{\dagger \,\,(m)}D_{E}^{(m)}\right) \nonumber\\
= i \left( D_{M}^{\dagger \,\,(m)}\otimes I^{(\otimes N)}\right)
D_{E}^{(m)} + i D_{E}^{\dagger \,\,(m)}\left( D_{M}^{(m)}\otimes
I^{(\otimes N)}\right)\label{unitary of T}
\end{eqnarray}
Now let us sandwich each term of (\ref{unitary of T}) with
$|m\rangle $. Since $\langle m | D_E^{(m)} | m\rangle =0$, then:
\begin{eqnarray}
\sum_{j , l_j}A_{E}^{\dagger \,\,(l_j,m_j)}A_{E}^{(l_j,m_j)}=-
\left( i D_{M}^{(m)} - i D_{M}^{\dagger \,\,(m)}+D_{M}^{\dagger
\,\,(m)}D_{M}^{(m)}\right) \label{unitarirelations}
\end{eqnarray}
This relation has a clear intuitive interpretation. With
the approximation (\ref{aproximacio de la matriu de scatering})
for the $T$ matrix, the particle cannot change the state of more
than one target at a time. Thus, for a given initial
configuration $(n,\{m\})$, the probability  to change target $j$
from state $m_j$ to state $l_j$ is:
\begin{eqnarray}
\parallel  \langle l_j \ | T^{(j)} \ \ | \phi^n
\rangle | m_j \rangle \parallel ^2
\end{eqnarray}
The probability that any target suffers any kind of change for
some initial particle state is, thus, written in terms of the $A_E$
operators:
\begin{eqnarray}
\sum_n p_n \sum_{j} \sum_{\{m_j\}}q_{m_j} \sum_{l_j \neq m_j}%
\langle \phi^n | A_{E}^{\dagger \,\,(l,m_{j})}A_{E}^{(l,m_{j})}|
\phi^n \rangle \label{prob any change}
\end{eqnarray}
The opposite situation, in which there is no change in the
slab, has a probability:
\begin{eqnarray}
\sum_{\{m\}}q_m \sum_n p_n \parallel \prod_j\langle m_j |%
(1+ i \sum_j T^{(j)}) \ \ | \phi^n \rangle \prod_j | m_j \rangle
\parallel^2 
\end{eqnarray}
that, written in terms of the $D_M$'s:
\begin{eqnarray}
1-\sum_n p_n\sum_{{m}}q_{m}  %
\langle \phi^n| i(D^{\dagger m}_M- D^{m}_M)-D^{\dagger m}_M
D^{m}_M|\phi^n \rangle  
\end{eqnarray}
These two probabilities must add up to one, which expresses the unitarity relation (\ref{unitarirelations}).

\section{The equation of motion}

In the previous section we have found an equation that relates the
density matrices for the particle before and after it crosses a
slab of matter, which we called the one-step equation. From it we
now derive the differential equation for the evolution of this
density matrix.

\subsection{From the one-step to a time equation}

\label{l'equacio en el temps}
Let us first justify the procedure we follow with a simple
example which also applies to the general case.

Suppose the particle is in a pure state and the medium is made out
of static scattering centers. After the particle crosses the \textit{r}th slab, the state for the particle is:
\begin{eqnarray}
|\phi(r)\rangle = (1 + i T)|\phi(r-1)\rangle \label{one step
simplified}
\end{eqnarray}
The particle spends a time $\Delta t=\frac{\delta}{v}$ crossing
the slab, where $\delta$ is its thickness and $v$ is the mean
velocity of the particle. Then, one could be na\"{\i}vely tempted to
write an equation of the sort:
\begin{eqnarray}
\frac{|\phi(r)\rangle -|\phi(r-1)\rangle}{\Delta t} = i
\frac{v}{\delta}T|\phi(r-1)\rangle \ \ \ \ \ \
\stackrel{\textrm{continuum}}{\line(1,0){80}\vector(1,0){10}} \ \
\ \ i \frac{d}{dt}|\phi\rangle = -\frac{v}{\delta}T |\phi\rangle,
\end{eqnarray}
But this cannot be a Schr\"odinger equation because $T$ is not an
hermitian operator. Demanding that $(1+iT)$ should be of the form
$exp(-i H_{eff} \Delta t)$, the correct differential
equation is:
\begin{eqnarray}
\frac{d}{dt}|\phi\rangle = \frac{v}{\delta}\textrm{ln}(1+iT)
|\phi\rangle \label{eq schro one step}
\end{eqnarray}
Unfortunately, these arguments cannot be immediately generalized
to the case of equation (\ref{one-step-eq}), because in that case
the particle gets entangled with the medium, and the one-step
equation is much more complicated than (\ref{one step
simplified}); moreover, we are aiming at an evolution equation of
the Lindblad type, which is not a mere Schr\"odinger evolution. For
that purpose lets go back to our example in (\ref{one step
simplified}) and re derive (\ref{eq schro one step}) in such a way
that now it does generalize for the \textit{the one step equation} (\ref{one-step-eq}).

Consider that the particle crosses $r$ slabs in a time $\Delta t=r
\frac{\delta}{v}$. We can iterate equation (\ref{one step
simplified}) $r$ times and expand $(1+ i T)^r$ :
\begin{eqnarray}
|\phi(r)\rangle = (1+r (iT)+\frac{r(r-1)}{2}( i T )^2+...)
|\phi(0)\rangle
\nonumber \\
 = (1+r(i T-\frac{(iT)^2}{2}+...)+O(r^2))|\phi(0)\rangle
\label{one step taylor en r}
\end{eqnarray}
Compare (\ref{one step taylor en r}) with the Taylor expansion of
$|\phi(t+\Delta t)\rangle$:
\begin{eqnarray}
|\phi(t+\Delta t)\rangle = |\phi(t)\rangle+\Delta
t\frac{d}{dt}|\phi(t)\rangle+... \label{one step taylor en t}
\end{eqnarray}
where  $|\phi(t+\Delta t)\rangle=|\phi(r)\rangle$ and
$|\phi(t)\rangle=|\phi(0)\rangle$. Now, identify the terms of
(\ref{one step taylor en r}) linear in $\Delta t$ with those of
(\ref{one step taylor en t}) through the relation $\Delta t=r
\frac{\delta}{v}$, and write:
\begin{eqnarray}
\frac{d}{dt}|\phi(t)\rangle = \frac{v}{\delta} ( i
T-\frac{(iT)^2}{2}+...)|\phi(t)\rangle
\end{eqnarray}
which is nothing but the Taylor expansion of $\textrm{ln}(1+i T)$.
Thus, we retrieve (\ref{eq schro one step}).

We are now ready to apply this procedure to the one-step equation
(\ref{one-step-eq}). To derive the desired relation let us Taylor
expand $\rho$ about the instant t, as in (\ref{one step taylor en
t}):
\begin{eqnarray}
\rho (t+\Delta t)\simeq \rho (t)+\Delta t~\dot{\rho}(t)+...
\end{eqnarray}
On the other hand we know $\rho $ after $r$  steps by iterating
equation (\ref{one-step-eq}) $r$ times. Expanding the result in
powers of $r$:
\begin{eqnarray}
\rho (r)\simeq \rho (0)+r~\Lambda\left[ \rho \right] +...
\label{l'operaddor L}
\end{eqnarray}
where $\Lambda$ is a linear operator acting on $\rho$. By comparing
equal powers of $\Delta t$ in the two expressions, one finally
gets:
\begin{eqnarray}
\dot{\rho}(t)=\frac{\textrm{v}}{\delta }\Lambda\left[ \rho \right]
\label{step & time}
\end{eqnarray}
This is the equation of motion, that, as we will see in the
following section, has a Lindblad form.

\subsection{The equation of motion for the reduced density matrix}

Here we compute $\Lambda[\rho]$ by iterating eq.(\ref{one-step-eq}),
as explained in the previous section. If we only keep terms linear
in the potential, the iteration only gives a linear term in
$r$:
\begin{eqnarray}
r\sum_{m}q_{m} \left( i D_{M}^{(m)}\,\rho \, - i
\rho~D_{M}^{\dagger \,\,(m)}\right),
\end{eqnarray}
and (\ref{step & time}) is an equation of Schr\"odinger type . It is not until we
retain terms up to quadratic  in the potential that we find any sign of decoherence and, hence, corrections to this standard evolution. These
terms are:
\begin{eqnarray}
& &r\sum_{m}q_{m} \left( %
i D_{M}^{(m)}\,\rho \, - i \rho~D_{M}^{\dagger
\,\,(m)}\,+D_{M}^{(m)}\,\rho D_{M}^{\dagger
\,\,(m)}\right) \nonumber\\
+& & r(r-1)\left( \sum_{m}q_{m}D_{M}^{(m)}\right) \rho\left(
\sum_{n}q_{n}D_{M}^{\dagger \,\,(n)}\right) \nonumber\\
+& &\frac{r(r-1)}{2}\sum_{m,n}q_{m}q_{n}\left(
-D_{M}^{(m)}D_{M}^{(n)}\rho - \rho D_{M}^{\dagger
\,\,(m)}D_{M}^{\dagger \,\,(n)}\right) \nonumber \\
+& &r \sum_{m_j, l_j}q_{m_j}  A_{E}^{(l_j,m_j)}\rho A_{E}^{\dagger
(l_j,m_j)}
\end{eqnarray}
from which we retain only the pieces linear in $r$. Defining the operators
$A_{M}^{n,m}\equiv D_{M}^{(m)}-D_{M}^{(n)}$, we can write the equation
of motion (\ref{step & time}) as:
\begin{eqnarray}
\dot{\rho}(t)= -i H \rho + i\rho H^{\dagger}
+\frac{1}{2}\frac{\textrm{v}}{\delta }\sum_{m,n}q_{m}q_{n}
A_{M}^{n,m} \rho A_{M}^{\dag n,m} +\frac{\textrm{v}}{\delta }
\sum_{l_j,m_j}q_{m_j} A_{E}^{(l_j,m_j)}\rho A_{E}^{\dagger
(l_j,m_j)}\label{eq en la forma index refraccio}
\\
 \mathrm{ Where  \ \ \ \ \ \ \ \      }   H\equiv -\frac{\textrm{v}}{\delta } \left(
\sum_{\{m\}} q_mD^{(m)}_M -i\frac{1}{2} (\sum_{\{m\}} q_mD^{(m)}_M)^2
\right) \ \ \ \ \
\nonumber
\end{eqnarray}
This equation is not in the Lindblad form yet. In order to do that, we use the unitarity relations of section (\ref{unitary relations}) and rewrite it as:
\begin{eqnarray}
\dot{\rho}=-i\left[\,H_{eff}\,,\,\rho \, \right]
+\frac{1}{4}\frac{\textrm{v}}{\delta
}\sum_{m,n}q_{m}q_{n}L^{(n,m)}_M [\rho]
+ \frac{1}{2}\frac{\textrm{v}}{\delta }\sum_{l_j,m_j}q_{m_j}
L^{(l_j,m_j)}_E [\rho] \label{equacio final matiudensitat}
\end{eqnarray}
This is the final expression. We have defined the effective hamiltonian and the  Lindblad
terms as:
\begin{eqnarray}
&& \! \! \!H_{eff}=-\frac{1}{2}\frac{\textrm{v}}{\delta }\left(
\sum_{\{m\}} q_mD^{(m)}_{M}+\sum_{\{m\}} q_mD_{M}^{\dagger (m)}\right) \label{hamiltonia efectiu} \\
&& \! \! \! L^{(n,m)}_M [\rho] =  2~A_{M}^{n,m}~\rho
~A_{M}^{\dagger \,\,\,n,m}-A_{M}^{\dagger
\,\,\,n,m}A_{M}^{n,m}~\rho
-\rho ~A_{M}^{\dagger \,\,\,n,m}A_{M}^{n,m} \nonumber \\
&& \! \! \! L^{(l_j,m_j)}_E [\rho]= 2~\,A_{E}^{(l_j,m_j)}\,\,\rho
~A_{E}^{\dagger (l_j,m_j)}-A_{E}^{\dagger
(l_j,m_j)}A_{E}^{(l_j,m_j)}\rho -\rho ~A_{E}^{\dagger
(l_j,m_j)}A_{E}^{(l_j,m_j)} \label{definicio Lm i Le}
\end{eqnarray}
As we can see, the hamiltonian is hermitian, and it contains the
"mean" effects that the medium does on the particle. To first
order in the potential it is nothing but the mean energy felt by
the particle while it propagates inside the medium.

An important remark concerning the Lindblad operators $A_M$ and
$A_E$ is in order. Mixture effects only depend upon how much mixed
the medium is: if it is pure $A_M=0$ and  they are not present.
However, footprint effects are unavoidable once the particle has
modified the state of the medium. This always carries $A_E\neq 0$,
regardless of whether the state is mixed or not. \label{seccio 5.2}

\subsection{The independence on $\delta$ of the equation of motion}

Physics cannot depend on $\delta$, the width of the elementary slab of our partition; any dependence on $\delta$ in equation (\ref{eq en la forma index refraccio}) is an artifact of the approximations that have been used and it will be automatically removed whenever the approximations hold. In order to qualify this statement, let us regard again, for simplicity, the case in section \ref{l'equacio en el temps}, where the particle always evolves as a pure state: we expect that the effective hamiltonian $H=\frac{v}{\delta}\ln(1+iT^{(\delta)})$ does not depend on the width of the slab, while $T^{(\delta)}$ does.

This follows from the relation between the $S$ matrix for a slab of width $\delta$ and the $S$ matrix for a slab twice as thick:
\begin{eqnarray}
(1+iT^{(\delta)})(1+iT^{(\delta)})=(1+iT^{(2\delta)}) \label{la relacio tonta}
\end{eqnarray}
from which we obtain
\begin{eqnarray}
\frac{v}{2\delta}\ln(1+iT^{(2\delta)})=\frac{v}{\delta}\ln(1+iT^{(\delta)})
\end{eqnarray}
This shows the independence of $H$ on the $\delta$ parameter.  If we make some approximation in the computation of $T^{(\delta)}$,  the relation (\ref{la relacio tonta}) will fail to hold and it will lead us to a  spurious dependence on $\delta$.  Nevertheless, this is not substantial in the domain where the approximation for $T$ is valid.

In our case, the approximation consists of taking as $T$ matrix the sum of the individual $T^{j}$ matrices of scattering of just one target $j$. In order for equation (\ref{la relacio tonta}) to be satisfied:
\begin{eqnarray}
(1+i\sum_{j\in slab 1}T^{j})(1+i\sum_{j \in slab 2}T^{j})=(1+i\sum_{j \in slab 1 + slab 2}T^{j}) \label{condicions invariança slab}
\end{eqnarray}
the crossed terms $T^{j}T^{i}$ should be negligible. These are similar conditions to those imposed by unitarity, put forward in section (\ref{aproximacions sobre el medi}). Recall that this approximation amounts to neglecting re-scattering, which implies that the density of targets must be small.

\section{The Lindblad equation for a gas}

In this section we treat the example of a gas. We expect in this
case some simplifications to occur in the general equation
(\ref{eq en la forma index refraccio}), because the gas is an
homogeneus medium. In particular, the simplifications apply to the
matrix elements $\langle k' |H|k \rangle$, that are related to the
refraction index of the gas, as we shall see in section 7.

\subsection{The hamiltonian part}

Let us consider the hamiltonian piece from (\ref{eq en la forma
index refraccio}) in momentum representation:
\begin{eqnarray}
\langle k | H |q \rangle \equiv  -\frac{\textrm{v}}{\delta }
\langle k |\sum_{\{m\}} q_m D_M^{(m)}|q \rangle
+i\frac{1}{2}\frac{\delta }{\textrm{v}}\langle k |\left(
\frac{\textrm{v}}{\delta } \sum_m q_m D_M^{(m)}\right)^2|q \rangle
\label{H dependent de delta}
\end{eqnarray}
In order to perform the computation of these sums, we divide the slab in boxes
of volume $1/n$ so as to have one target inside each box ($n$ is
the density of targets). We  also write $|m_{j}, x_j\rangle$ to
label explicitly the centers $x_j$ of the target wave functions,
and similarly for the weights $q_{m_j,x_j}$. The sum over all
possible locations $x_j$ inside $box_j$ becomes an integral:
\begin{eqnarray}
\langle k|\sum_{\{m\}} q_m D_M^{(m)}|q\rangle=\sum_j
\sum_{\{m_j,x_j\}}
q_{m_j,x_j}\langle k| D_M^{m_j,x_j} |q\rangle \nonumber \\
=\sum_j\sum_{\{m_j\}} q_{m_j} \int_{box_j} n dx_j
e^{i(q-k)x_j}\langle k|\langle m_j ,0 |T|m_j ,0\rangle|q\rangle
\end{eqnarray}
We have used the fact that the matrix element $\langle k|\langle
m_{j},x_j|T|m_{j},x_j \rangle|q\rangle = e^{i(q-k)x_j}\langle
k|\langle m_j  |T|m_j \rangle|q\rangle$, where it is understood
that the states $|m_j\rangle$ are centered at the origin now.
Since the weights $q_{m_j}$ are the same for all targets, the sum
over $j$, together with the integral over each box, becomes an
integral over the slab. Then,
\begin{eqnarray}
\langle k| \sum_{\{m_j\}} q_{m_j}D^{m_j}_M|q\rangle= n
\sum_{\{m\}} q_{m} \int_{slab} dx e^{i(q-k)x}\langle k|\langle m
|T|m \rangle|q\rangle\label{expressio per d tilde}
\end{eqnarray}
The integral over the slab can be split into an integral over its
width (z coordinate) times an integral over the plane of the slab.
We first perform  the integral over the plane, and find that the
result is proportional to the delta function of $(q-k)_{\|}$, i.e.,
the components parallel to the plane:
\begin{eqnarray}
\frac{\textrm{v}}{\delta}\langle k|\sum_{\{m_j\}} q_{m_j}
D^{m_j}_{M}|q\rangle = \frac{\textrm{v}}{\delta} (2\pi)^2 n \sum_m
q_m
\int_0^{\delta} dz%
\delta_{\|} (q-k)e^{i(q_z-k_z) z}\langle m|T(k,q)
|m\rangle\label{els elements de matriu de D tilde}
\end{eqnarray}

We compute the projection of the $T$-matrix in the center of mass
coordinates: $k_{cm}=k_{1}+k_{2}$ and
$k_{r}=\frac{m_{2}}{m_{t}}k_{1}-\frac{m_{1}}{m_{t}}k_{2}$, where
$1$ labels the particle and $2$, the target. In such coordinates
$T$ reads:
\begin{eqnarray}
\langle k_r|\langle k_{cm}|T|q_{cm}\rangle|q_r\rangle=-2\pi \delta
(k_{cm} -q_{cm})\delta (E_{q_r}-E_{k_r}) T_E (k_r,q_r;m_r);
\end{eqnarray}
here $m_r$ is the reduced mass. The combination of these two delta
functions, together with $\delta_{\|} (q-k)$ in (\ref{els elements
de matriu de D tilde}), gives:
\begin{eqnarray}
&&\delta (q_1-k_1) \delta(q_2 -k_2) \frac{1}{|\frac{k_{z,1}}{m_1}-\frac{k_{z,2}}{m_2}|}\label{les tres deltes}\\
+ &&\delta_{\|}  (q_1-k_1) \delta_z\left(
k_1+q_1-\frac{m_1}{m_2}(k_2+q_2)\right)  \delta_{\|} (q_2 -k_2)
\delta_z\left( k_2-q_2-2q_1+\frac{m_1}{m_2}(k_2+q_2)\right)
\frac{2m_1}{|k_1-q_1|}\nonumber
\end{eqnarray}
The delta functions in the two terms provide the energy momentum
conservation laws corresponding to one-dimensional elastic
scattering. The first term is diagonal in both particle and
target momenta; it corresponds to the process where each velocity
remains unchanged and gives the dominant contribution to (\ref{els
elements de matriu de D tilde}), because the  imaginary argument
of the exponential vanishes. The second term corresponds to the
solution where the relative velocity flips its sign, and gives a
negligible contribution  if the particle wave packet is
entirely contained inside the slab of width $\delta$: therefore
$k_z \delta \gg 1$ and the oscillations of the exponential factor
kill the integral over $z$. Then (\ref{els elements de matriu de D
tilde}) reads\footnote{Recall that $\frac{k_z}{m_1}=v$, since we assume that particle state is peaket around velocity $v$ pointing in $z$ direction}:
\begin{eqnarray}
\frac{\textrm{v}}{\delta}\langle k_1| \sum_{\{m_j\}}
q_{m_j}D^{m_j}_{M}|q_1\rangle =- (2\pi)^3 n \sum_m q_m \langle m|
\frac{1}{|1-\frac{m_1}{m_2}\frac{k_{z,2}}{k_{z,1}}|}T_E (k_r, k_r;
m_r) |m\rangle \delta(k_1-q_1)\label{no tinc intuicions al
respecte}
\end{eqnarray}

The hamiltonian (\ref{H dependent de delta}) contains, also, the square term $(\sum_{m_j}
q_{m_j}D^{m_j}_{M})^2$. We discard this piece, because it
is quadratic in $n$, which should be small by consistency of the approximation (see (\ref{condicions invariança slab})).
Finally, we get for the $H$ matrix elements:
\begin{eqnarray}
\langle k_1 | H |q_1 \rangle  = (2\pi)^3 n \sum_m q_m \langle m|
\frac{1}{|1-\frac{m_1}{m_2}\frac{k_{z,2}}{k_{z,1}}|}T_E (k_r, k_r;
m_r) |m\rangle \delta(k_1-q_1)\label{hamiltonia final}
\end{eqnarray}
We assume that the tipical velocity of a target is much smaller
than the particle velocity, and therefore  the denominator
$|1-\frac{m_1}{m_2}\frac{k_{z,2}}{k_{z,1}}|\sim 1$, regardless of
whether  $\frac{m_1}{m_2}$ is small or not.

This is our final expression for the hamiltonian of the particle. As we can check, there is no-dependence
on the width $\delta$ of the slab  left over in (\ref{hamiltonia final}). Further simplifications occur
in the limits of both  $m_1 \ll m_2$ and  $m_1 \gg m_2$, that we
now consider.

\vspace{0.7cm}

\noindent \textbf{The limit of a heavy target. }

\vspace{0.7cm}

\noindent Let us expand (\ref{no tinc intuicions al respecte}) in
powers of $\frac{m_1}{m_2} \ll 1$. The zeroth order term is:
\begin{eqnarray}
\langle k_1 | H |q_1 \rangle = (2\pi)^3 n \delta(k_1-q_1) T_E
(k_1,k_1;m_1) +O\left( \frac{m_1}{m_2} \right),\label{pitufo}
\end{eqnarray}
which is independent on the state of the targets.

There are two sources of corrections to this:  the denominator of
(\ref{no tinc intuicions al respecte}), and the $T_E$ matrix
elements through their dependence on $k_r$ and $m_r$.  In a gas,
the target momentum expectation value vanishes, i.e., $\sum_m
q_m\langle m|k_2|m\rangle=0$, and the corrections coming from the
denominator cancel. The first $\frac{m_1}{m_2}$ correction, thus,
reads:
\begin{eqnarray}
-\delta(k_1-q_1)  (2\pi)^3 n\frac{m_1}{m_2}k_1
\frac{\partial}{\partial k_1}T_E (k_1,k_1;m_r),
\end{eqnarray}
where we have not expanded $m_r$ in $T_E$. Notice that these
corrections start quadratic in the potential  because the linear
term in $V$ does not depend on $k_r$ at all. The final expression
in this limit reads:
\begin{eqnarray}
\langle k_1 | H |q_1 \rangle = \delta(k_1-q_1) (2\pi)^3 n \left( 1
- \frac{m_1}{m_2}(k_1 \frac{\partial}{\partial k_1}) \right) T_E
(k_1,k_1;m_r) +O\left( (\frac{m_1}{m_2})^2 \right)\label{caca de
vaca}
\end{eqnarray}
The corrections to the infinitely heavy target limit are
$O(\frac{m_1}{m_2},V^2)$.

\vspace{0.7cm}

\noindent \textbf{The limit of a heavy particle. }

\vspace{0.7cm}

\noindent Let us now consider the opposite limit, and expand
(\ref{no tinc intuicions al respecte}) in powers of
$\frac{m_2}{m_1} \ll 1$. The zeroth order term is:
\begin{eqnarray}
\langle k_1 | H |q_1 \rangle = (2\pi)^3 n \sum_m q_m \langle m|T_E
\left( \frac{m_2}{m_1}k_1- k_2, \frac{m_2}{m_1}k_1- k_2; m_2\right) |m\rangle \delta(k_1-q_1) +O\left( \frac{m_2}{m_1}
\right)
\end{eqnarray}
$\frac{m_2}{m_1}k_1= m_2 v_1$ should not be treated as a small quantity as compared to $k_2$ if we assume that the particle velocity $v_1$ is larger than the velocities of the targets in the gas. 

The first order corrections contain a term coming from the
denominator of (\ref{no tinc intuicions al respecte}) which does
not cancel as it did before. Finally, the hamiltonian $\langle k_1| H |q_1 \rangle $ reads:
\begin{eqnarray}
(2\pi)^3 n \sum_m q_m \langle m|
(1+\frac{m_2}{m_1}(\frac{k_1 k_2}{k_1^2}+k_2 \frac{\partial}{\partial K}))T_E (K=\frac{m_2}{m_1}k_1- k_, K;
m_r) |m\rangle \delta(k_1-q_1)+O\left( (\frac{m_2}{m_1})^2 \right)
\end{eqnarray}
where we have substitute the quotient $k_{z,2} / k_{z,1}$ by $k_1 k_2 / k_1^2$: they both coincide in the limit of a particle state pecked in the $z$ direction.

\subsection{The piece of decoherence}

Apart from $H$, the evolution equation (\ref{eq en la forma index
refraccio}) involves further pieces that contain the $A_M$ and
$A_E$ operators, which are the ones that induce decoherence. Let
us first consider the terms containing $A_M$ (see section
(\ref{seccio 5.2}) for details ):
\begin{eqnarray}
A_M \rho A_M^{\dagger} \equiv \frac{1}{2}\frac{\textrm{v}}{\delta
}\sum_{m,n}q_{m}q_{n} A_{M}^{n,m} \rho A_{M}^{\dag n,m} =
\frac{\textrm{v}}{\delta }\sum_{m}q_{m}D_{M}^{m} \rho D_{M}^{\dag
m} - \frac{\textrm{v}}{\delta }\sum_{m}q_{m} D_{M}^{m} \rho
\sum_{n} q_{n}D_{M}^{\dag n}\label{gamarus}
\end{eqnarray}
Making explicit the summation over all targets in (\ref{gamarus})
we find:
\begin{eqnarray}
A_M \rho A_M^{\dagger} = \frac{\textrm{v}}{\delta }\sum_j \sum_{m_j, x_j}
(q_{m_j,x_j}-q_{m_j,x_j}^2)D_{M}^{m_j,x_j} \rho D_{M}^{\dag
m_j,x_j}\label{suma discreta}
\end{eqnarray}
where we have split label $(m_j)$ into $(m_j,x_j)$ as before, to
explicitly display the center of target $j$ wave function. We
proceed as before and convert the sum over  $x_j$  into an
integral (note that the squared probability in (\ref{suma
discreta}) vanishes). We get:
\begin{eqnarray}
A_M \rho A_M^{\dagger} = n \frac{\textrm{v}}{\delta} \sum_{m} q_{m}
\int_{slab} d^3x D_{M}^{m,x} \rho D_{M}^{\dag m,x}
\end{eqnarray}

The term coming from the $A_E$ operator will have a similar
expression, yielding:
\begin{eqnarray}
A_E \rho A_E^{\dagger} \equiv \frac{\textrm{v}}{\delta } \sum_{m_j , l_j\neq
m_j}q_{m_j} A_{E}^{(l_j,m_j)}\rho A_{E}^{\dagger (l_j,m_j)}=n
\frac{\textrm{v}}{\delta} \sum_{m} q_{m}\sum_{l\neq m} \int_{slab}
d^3x A_{E}^{l,m,x} \rho A_{E}^{\dag l, m,x}
\end{eqnarray}
 The matrix elements $(k',k)$ of these two terms together add up to:
\begin{eqnarray}
\langle k^{\prime}|A_M\rho A_M^{\dagger}+A_E\rho A_E^{\dagger} |k\rangle  = &&  n
\frac{\textrm{v}}{\delta} \sum_{l,m} q_{m} \int  dq dq^{\prime
}\,(2\pi)^2 \delta _{\parallel }\left( k^{\prime
}+q-k-q^{\prime }\right) \times \\
&\times\!&\int _{0}^{\delta }dz\, e^{i\left( \left(
k_z^{\prime}-k_z\right) -\left( q_z^{\prime }-q_z\right) \right)
z} \langle k^{\prime}| \langle l |T | m\rangle |q^{\prime }\rangle
\rho \left( q^{\prime },q\right) \langle q|\langle
m|T^{\dagger}|l\rangle |k\rangle \nonumber \label{els elements de matriu de
la part decoherent}
\end{eqnarray}
The T-matrix elements involved in the last expression, to first
order in the potential, and in momentum representation, are:
\begin{eqnarray}
\langle k'_1 |\langle l |T | m\rangle|q_1'\rangle= -2\pi\int \int
d^3k'_{2} \ d^3q'_{2} A_{l}^{*}(k'_{2}) \delta (k'_{cm}-q'_{cm})
\delta (E_{q'_r}-E_{k'_r})
\widetilde{V}(q_{r}-k_{r})A_{m}(q_{2}) \nonumber \\
\langle k_1 |\langle l |T^{\dagger} | m\rangle|q_1\rangle=
-2\pi\int \int d^3k_{2} \ d^3q_{2} A_{l}(k_{2}) \delta
(k_{cm}-q_{cm}) \delta (E_{q_r}-E_{k_r})
\widetilde{V}^{*}(q_{r}-k_{r})A^{*}_{m}(q_{2}) \label{computation
de l'exemple: plantejament}
\end{eqnarray}
where $A_l$, $A_m$ are the target wave functions. A resolution of
the identity $\sum_l
A_{l}^{*}(k'_{2})A_{l}(k_{2})=\delta(k_2-k'_2)$  appears in
(\ref{els elements de matriu de la part decoherent}) once
expressions (\ref{computation de l'exemple: plantejament}) are
used.
\begin{eqnarray}
\langle k^{\prime}|A_M\rho A_M\! \! &+ & \! \!A_E\rho A_E|k\rangle
=   n \frac{\textrm{v}}{\delta}(2\pi)^4 \sum_{l,m} q_{m} \int
d^3k_2 d^3q_1 d^3q_2 d^3k'_2 d^3q'_1 d^3q'_2  \int _{0}^{\delta }
dz\,
e^{i\left( \left( k_z^{\prime}-k_z\right) -\left( q_z^{\prime }-q_z\right) \right) z}\nonumber \\
&&\times \delta(k_2-k'_2) \delta _{\parallel }\left( k_1^{\prime
}+q_1-k_1-q_1^{\prime }\right) \delta (k_{cm}-q_{cm}) \delta (k'_{cm}-q'_{cm}) \delta(E_{q'_r}-E_{k'_r})\delta (E_{q_r}-E_{k_r}) \nonumber  \\
&&\times
A_{m}(q_{2})A^{*}_{m}(q'_{2})\widetilde{V}^{*}(q_{r}-k_{r})
\widetilde{V}(q'_{r}-k'_{r}) \rho(q'_1,q_1)\label{expressio sense
pertorvar}
\end{eqnarray}
This is our final expression for Lindblad piece of the equation of motion (\ref{eq en la forma index refraccio}).

Further simplifications occur, again, in the limits of  heavy
target and heavy particle. The essential difference between the
two limits enters through the energy delta functions.

In both cases, the following relation is useful to write a perturbative expansion in the ratio of masses:
\begin{eqnarray}
\delta (E_{q_r}-E_{k_r}) =\frac{1}{2 \pi i}
\left(\frac{1}{E_{q_{r}}-E_{k_{r}}- i \epsilon}-
\frac{1}{E_{q_{r}}-E_{k_{r}}+i \epsilon} \right) \label{la
representacio de la delta}
\end{eqnarray}
We now study the two different limits separately.

\vspace{0.7cm}

\noindent \textbf{The limit of a heavy target. }

\vspace{0.7cm}

\noindent In such limit we can expand the terms  in (\ref{la
representacio de la delta}) as a geometric  series:
\begin{eqnarray}
\frac{1}{E_{q_r}-E_{k_r} \mp i \epsilon}= \frac{2m_r}{q_1^2-k^2_1
\mp i \epsilon}Ê \sum_{s=0}^{\infty}
\left(\frac{1}{2}\frac{m_1}{m_t}\frac{(q_1+k_2)^2-(k_1+q_2)^2}{q_1^2-k^2_1
\mp i \epsilon}\right)^s 
\end{eqnarray}
with the caution that with this expansion the series thus obtained
is asymptotic. Here we have used  the conservation of the center
of mass momentum $\delta (k_{cm}-q_{cm})$ present in
(\ref{expressio sense pertorvar}). By taking higher derivatives of
(\ref{la representacio de la delta}) we have
\begin{eqnarray}
\frac{1}{2 \pi i}\left[ \frac{1}{\left( E_{q_1}-E_{k_1} - i
\epsilon \right)^{s+1}}-\frac{1}{\left(E_{q_1}-E_{k_1} + i
\epsilon\right)^{s+1}}\right]=
\frac{(-1)^s}{s!} \frac{\partial^s \delta (E_{q_1}-E_{k_1})}{\partial 
E_{q_1}^s} \label{pertorbacio denominador}
\end{eqnarray}
that allows us to write (\ref{la representacio de la delta}) as:
\begin{eqnarray}
\delta (E_{q_r}-E_{k_r})= \sum_{s=0}^{\infty} \frac{1}{s!} (\frac{-1}{4m_2})^s\frac{\partial^s \delta (E_{q_1}-E_{k_1})}{\partial
E_{q_1}^s}
\left(  (q_1+k_2)^2-(q_2+k_1)^2 \right)^s
\end{eqnarray}
suitable  for an expansion in inverse powers  of the target mass. Let us focus on the zeroth order term of this expansion. The set of delta functions in (\ref{expressio sense
pertorvar}) is now:
\begin{eqnarray}
 \delta(k_2-k'_2) \delta _{\parallel }\left( k_1^{\prime}+q_1-k_1-q_1^{\prime }\right)
 \delta (k_{cm}-q_{cm}) \delta (k'_{cm}-q'_{cm})
\delta(E_{q'_1}-E_{k'_1})\delta (E_{q_1}-E_{k_1})\label{la inspiracio}
\end{eqnarray}
At this point, we replace such string of deltas by the following expression
\begin{eqnarray}
\frac{1}{|v_{z,1}|} \delta(k_2-k'_2) \delta \left(
k_1^{\prime}+q_1-k_1-q_1^{\prime }\right)
 \delta (k_{cm}-q_{cm}) \delta (k'_{cm}-q'_{cm})
\delta(E_{q_1}-E_{k_1})\label{generalitzacio}
\end{eqnarray}
Let us justify this substitution. In first place, the two terms
have the same trace in the limit where the state for the particle
is peaked around a velocity $v$ and the size of the wave packet
is smaller than $\delta$, the width of the slab. Also, in such
limit (where $\rho(q,q')$ is different from zero only if $q \approx
q'$), the regions of momenta $q,q',k,k'$ for which the arguments of the
delta functions vanish are the same in both expressions. In this limit of $q\approx q'$, from  (\ref{la inspiracio}) we find in principle two allowed regions:
\begin{eqnarray}
(k'-k)_{\|}=(q-q')_{\|} \approx 0 \ ; \ E_{k'}=E_{q'}\approx E_q=E_k  
\ \ \ \ \rightarrow\ \ \ \  \left( k\sim k'  \right) \textrm{ or }  \left( k'_{\|} \sim k_{\|}
\ ,\ k_z \sim -k'_z \right)
\end{eqnarray}
but the only one that survives upon $z$ integration in (\ref{expressio sense
pertorvar}), is the first one, with $k\sim k'$. This is the
same region that makes the arguments of
the delta functions in (\ref{generalitzacio}) vanish too.
\vspace{0.5 cm}

\noindent The zeroth order term in $1/m_2$ of (\ref{expressio
sense pertorvar}) is, thus:
\begin{eqnarray}
&&\langle k^{\prime}|A_M\rho A_M+A_E\rho A_E|k\rangle  = \\
 &&n(2\pi)^4 \int dq_1 dq'_1 \delta \left(
k_1^{\prime}+q_1-k_1-q_1^{\prime }\right)
\delta(E_{q_1}-E_{k_1})
\widetilde{V}^{*}(q_{1}-k_{1})
\widetilde{V}(q'_{1}-k'_{1})\rho(q'_1,q_1)
\nonumber
\end{eqnarray}
Notice that this is totally independent on the specific form of the target states and preserves the mean value of the particle energy.

\vspace{0.7cm}

\noindent \textbf{The limit of a heavy particle. }

\vspace{0.7cm}

\noindent Let us now consider the opposite limit of a heavy particle. The terms in (\ref{la representacio de la delta}) can  also be expanded as a geometric series, now in powers of $\frac{m_2}{m_1}$. The relative energy $E_{q_r}$ is:
\begin{eqnarray}
E_{q_r}=\frac{1}{2m_r}(-q_2+\frac{m_2}{m_t}q_1+\frac{m_2}{m_t}q_2)^2
\end{eqnarray}
As in the hamiltonian piece, we cannot treat $\frac{m_2}{m_t}q_1$ as small quantity in front of $q_2$, since  we assume that the particle velocity $v_1$ is larger than the velocities of the targets in the gas. 

We rewrite $E_{q_r}$ as:
\begin{eqnarray}
E_{q_r}=\frac{1}{2m_r}(-q_2+m_r v_1 +\frac{m_2}{m_t}(\Delta q_1+q_2))^2
\end{eqnarray}
where $\Delta q_1= q_1-m_1 v_1$. This brings our expressions to a form that can be easily expanded in powers of $\frac{m_2}{m_t}(\Delta q_1+q_2)$, the fluctuations of the particle and target momenta about their approximate mean values.   (\ref{la representacio de la delta}) reads:
\begin{eqnarray}
\frac{1}{E_{q_r}-E_{k_r} \mp i \epsilon}= 
\frac{2m_r}{(m_r v_1 -q_2)^2-(m_r v_1 -k_2)^2 \mp i \epsilon}Ê
\sum_{s=0}^{\infty}
\left(\frac{2m_2}{m_t}\frac{(k_2-q_2)(\Delta q_1+q_2)}{(m_r v_1 -q_2)^2-(m_r v_1 -k_2)^2 \mp i \epsilon}\right)^s 
\end{eqnarray} 
that is an  expansion  in inverse powers of the particle mass.  

To zeroth order, the delta functions that appear in (\ref{expressio sense
pertorvar}) are now:
\small
\begin{eqnarray}
\delta(k_2-k'_2) \delta_{\parallel}(k'_1+q_1-k_1-q'_1) \delta (k_{cm}-q_{cm})\delta (k'_{cm}-q'_{cm})
\delta(E_{m_2 v_1-q_2}-E_{m_2 v_1-q'_2}) \delta(E_{m_2 v_1-k_2}-E_{m_2 v_1-q_2})
\end{eqnarray}
\normalsize
At this point we use the fact that the velocity of the particle is much bigger than the target velocity, and the last expression reads:
\begin{eqnarray}
\frac{1}{|v_{z,1}|}
\delta(k_2-k'_2) \delta(q_2-q'_2)
\delta(k'_1+q_1-k_1-q'_1) \delta (k_1+k_2-q_1-q_2)
\delta(E_{m_2 v_1-k_2}-E_{m_2 v_1-q_2})
\end{eqnarray}
Finally, equation (\ref{expressio sense pertorvar}) is, to zeroth order:
\begin{eqnarray}
&&\langle k^{\prime}|A_M\rho A^{\dagger}_M\! \! +  \! \!A_E\rho A^{\dagger}_E|k\rangle
=  n (2\pi)^4  \sum_{m} q_m \int d^3k_2 d^3q_2 d^3q'_1 d^3q_1
\\
&&%
\delta(k'_1+q_1-k_1-q'_1) \delta (k_1+k_2-q_1-q_2)
\delta(E_{m_2 v_1-k_2}-E_{m_2 v_1-q_2})
\rho(q'_1,q_1)
|A_{m}(q_{2})|^2 |\widetilde{V}^{*}(k_{2}-q_{2})|^2 \nonumber
\end{eqnarray}

\section{The index of refraction}

The index of refraction is usually defined as the phase shift that
the medium induces on a plane wave as the particle propagates
through the medium. In an evolution where coherence is not
preserved, this concept, as it stands, does not hold anymore. This
is the case of the Lindblad evolution. Yet, a generalization is
still possible for the part of $\rho$ that preserves coherence. Let us first identify such part regarding the way a Lindblad equation increases the mixing of the density matrix.
A Lindblad evolution is of the form:
\begin{eqnarray}
\dot{\rho}=-iH\rho +i\rho H^{\dagger}+2A \rho
A^{\dagger},\label{entenent lindblad}
\end{eqnarray}
where $H$ is $H_{eff} -iA^{\dagger}A$. After an infinitesimal time
interval, the density matrix evolves to:
\begin{eqnarray}
\rho(t+\delta t)=\rho(t) + i \delta t (H \rho - \rho
H^{\dagger})+\delta t ( 2 A \rho A^{\dagger}) \label{hostia}
\end{eqnarray}
The first two terms together add up to a density matrix which has
trace less than one (since $H$ is non-hermitian) and coherence is
preserved on it (since it is an hamiltonian evolution). There is a
further term, the last one in (\ref{hostia}), which is also a
density matrix on its own. Thus, we see that under such a time
evolution a density matrix can always be split as a part in which
coherence is preserved, $\rho^{coh}$, plus the rest, $\rho^{mix}$,
which is also a density matrix (neither one is normalized here). This mechanism reflects the fact that the average of two density matrices is, in general, a density matrix with more
mixing. This suggests a definition of  $\rho^{coh}$ so that it satisfies the hamiltonian equation:
\begin{eqnarray}
\frac{\partial}{\partial t}\rho^{coh}= -i H\rho^{coh}+ i
\rho^{coh} H^{\dagger}; \label{ro coherent}
\end{eqnarray}
then, for $\rho^{mix}$:
\begin{eqnarray}
\frac{\partial}{\partial t}\rho^{mix}= -iH\rho^{mix} +i\rho^{mix}
H^{\dagger}+2A \rho^{mix} A^{\dagger}+2A \rho^{coh}
A^{\dagger},\label{eq matreiu mix}
\end{eqnarray}
which is the same as (\ref{entenent lindblad}) plus a term that
depends on $\rho^{coh}$. This non-homogeneous term is the
responsible for the increasing of the trace of $\rho^{mix}$, in order to maintain properly normalized the total density matrix $\rho$. Equation
(\ref{eq matreiu mix}) also ensures that $\rho^{mix}$ is a positive
matrix, since its time derivative  is a sum of positive matrices.

It is natural, thus, to write $\rho$ in equation (\ref{eq en
la forma index refraccio}) as $\rho^{coh}+ \rho^{mix}$, with
$\rho^{coh}$ and $\rho^{mix}$ satisfying the equations:
\begin{eqnarray}
\frac{\partial}{\partial t}\rho^{coh}&=& -i
\left[H\rho^{coh} - \rho^{coh} H^{\dagger}
\right] \ \ \ \ \ \  H\equiv -\frac{\textrm{v}}{\delta } \left(
\sum_{\{m\}} q_mD^{(m)}_M -i\frac{1}{2} (\sum_{\{m\}} q_mD^{(m)}_M)^2
\right) \label{definicio h coherent}\\
\frac{\partial}{\partial t}\rho^{mix}(t)&=&-i\left[\,H_{eff}\,,\,\rho^{mix} \, \right]
+\frac{1}{4}\frac{\textrm{v}}{\delta
}\sum_{m,n}q_{m}q_{n}L^{(n,m)}_M [\rho^{mix}]
+ \frac{1}{2}\frac{\textrm{v}}{\delta }\sum_{l_j,m_j}q_{m_j}
L^{(l_j,m_j)}_E [\rho^{mix}] \nonumber \\
&+&\frac{1}{2}\frac{\textrm{v}}{\delta }\sum_{m,n}q_{m}q_{n}
A_{M}^{n,m} \rho^{coh} A_{M}^{\dag n,m} +\frac{\textrm{v}}{\delta
} \sum_{l_j,m_j}q_{m_j} A_{E}^{(l_j,m_j)}\rho^{coh} A_{E}^{\dagger
(l_j,m_j)} \label{ro mix}
\end{eqnarray}
We naturally take $\rho^{coh}(0)=\rho(0)$ and $\rho ^{mix}(0)=0$ as initial conditions.

In spite of the fact that  $\rho$ decoheres, an index of refraction can be defined from $\rho^{coh}$, since it still evolves with an hamiltonian, which  has  already been computed for the case of an homogeneous gas in terms of the forward scattering amplitude. In order to retrieve the refraction index we should add the free hamiltonian, and get:
\begin{eqnarray}
\frac{k'}{k}=\sqrt{1- \frac{ \langle k|H|k\rangle}{k^2 / 2m_1}}
\end{eqnarray}
In the limit of heavy targets,  $\frac{m_1}{m_2} \ll 1$, we obtain:
\begin{eqnarray}
\frac{k'}{k}\approx
1+2\pi n
\frac{m_1}{m_r}\frac{(1-\frac{m_1}{m_2}k \frac{\partial}{\partial k}))f(k,k,m_r)}{k^{2}}+O\left( (\frac{m_1}{m_2})^2 \right)
\label{index of refraction}
\end{eqnarray}
This is the generalization of the result obtained by Fermi in the case of scattering centers \cite{Fermi}. Notice that the imaginary part of  (\ref{index of refraction}) has a clear interpretation: it is the depletion of the amplitude modulus due to a loss of coherence by dispersion.

The opposite situation of heavy particle, when $\frac{m_1}{m_2} \gg 1$, has also been computed in the previous section, with the resulting index of refraction:
\begin{eqnarray}
\frac{k'}{k} \approx 1+ 2 \pi n\frac{m_1}{m_r}  \frac{C}{k^2}
\end{eqnarray}
where the function $C$ is:
\begin{eqnarray}
C= \sum_m q_m \langle m| \left(1+\frac{m_2}{m_1} (k_1 k_2 / k_1^2 +  k_2 \frac{\partial}{\partial K} )\right)f(K=\frac{m_2}{m_1} k_1-k_2,K; m_r) |m\rangle
\end{eqnarray}

\section{Decoherence effects and interference patterns}

The decomposition found in the previous section allows us to see how interference patterns are destroyed by the decoherence induced through the interaction with the medium. Such patterns appear by the interference of  two states, $|\phi\rangle$  and $|\psi\rangle$, that travel following different paths and rejoin later. Thus, a superposition of $|\phi\rangle$ and $|\psi\rangle$  must retain enough coherence so that these interference patterns can be observed. For the  rest of this section we assume that the initial state $|\Phi\rangle$ is the sum  $|\phi\rangle + |\psi\rangle$. We thus start out with $\rho^{coh}(0)=|\Phi \rangle \langle \Phi|$ and $\rho^{mix}(0)=0$.

The crossed terms $|\phi \rangle \langle \psi|$ of the $\rho$ matrix are the responsible for the interferences: we argue that in the case when $|\phi\rangle$  and $|\psi\rangle$ are well separated and localized they decay exponentially in $\rho(t)$. The proof is in two steps. We first recall that the whole $\rho^{coh}$ decays in time, straightforwardly from its construction, in particular its crossed terms.  Then we see that these crossed terms never feed-back into $\rho$ through $\rho^{mix}$ via eq. (\ref{eq matreiu mix}) if the two states are separated enough and localized.

The proof goes by integrating the differential equation (\ref{eq matreiu mix}) for $\rho^{mix}$  for an infinitesimal time  interval $\delta t$:
\begin{eqnarray}
\rho^{mix}(\delta t)= \delta t \frac{1}{2}\frac{\textrm{v}}{\delta }\sum_{m,n}q_{m}q_{n}
A_{M}^{n,m} \rho^{coh}(0) A_{M}^{\dag n,m} +\delta t \frac{\textrm{v}}{\delta
} \sum_{l_j,m_j}q_{m_j} A_{E}^{(l_j,m_j)}\rho^{coh}(0) A_{E}^{\dagger
(l_j,m_j)}\label{termes coherent a ro mix}
\end{eqnarray}
All crossed terms $|\phi \rangle \langle \psi|$ appearing on the r.h.s. of this equation vanish.  The crossed terms in the second piece are directly zero, because $A_{E}^{(l_j,m_j)} $ is the amplitude for target $j$ to jump from $|m_j\rangle$ to $|l_j\rangle$ by interacting with the particle: when target $j$ is
close to the state $|\phi \rangle$ it is far from the state $|\psi \rangle$,
then $A_{E}^{(l_j,m_j)}  |\psi \rangle=0$, and vice-versa. As for the crossed terms from the first piece, they are proportional to:
\begin{eqnarray}
\sum_{\{m,n\}}\left( \prod_l q_{m_l} \prod_{l'}q_{n_{l'}}\right)
\left( \sum_j(D_M^{m_j}-D_M^{n_j}) \right) |\phi \rangle
\langle \psi| \left( \sum_i(D_M^{\dagger m_i}-D_M^{\dagger n_i})\right)\label{aquest arg es brutal}
\end{eqnarray}
and since states $|\phi\rangle$ and $|\psi\rangle$ are far away, targets that contribute in the summation over the $j$ index are different from those that contribute in the summation over $i$ index. This allows us to write (\ref{aquest arg es brutal}) as:
\begin{eqnarray}
 \left(\sum_{\{\textrm{j close to }\phi\}}\sum_{m_j,n_j}q_{m_j}q_{n_j}(D_M^{m_j}-D_M^{n_j})\right)
|\phi \rangle \langle\psi|
\left(\sum_{\{\textrm{i close to }\psi\}}\sum_{m_i,n_i}q_{m_i}q_{n_i}(D_M^{\dagger m_i}-D_M^{\dagger n_i})\right)=0
\end{eqnarray}
Then, after an infinitesimal time $\delta t$, crossed terms stay out from $\rho^{mix}(t)$. This completes the proof. 

We apply this result to the study of the interference patterns in a two slit Young  experiment. We shall study separately the probability distribution of hits on the screen that comes out from $\rho^{mix}$, and from  $\rho^{coh}$, each one being a density matrix on its own (except for an unessential normalization factor). Finally, the observed pattern is retrieved as the sum of both.

The part of $\rho^{coh}$ is easy: It is essentially the same as if there were no medium, with the wave number corrected $k \to k'$. This changes the oscillation wave-length to $\frac{2\pi}{ Re(k')}\frac{L}{D}$ and introduces a global damping factor of  $e^{-2Im(k')L}$ which diminishes the amplitude on the screen by this amount. 

Unlike $\rho^{coh}$, the part of $\rho^{mix}$ is far more involved to obtain, for it contains all the decoherence effects and it is the one which eventually washes out the interferences when the propagation of the particle between the slits and the screen is through a medium, rather than the vacuum. However, it is easy to see that it produces an approximately constant background in the region close to the central peaks. 

The argument is indirect and uses the result of an auxiliary problem, that  has as initial state $\widetilde{\rho}(0)=\frac{1}{2}|\phi \rangle \langle \phi|+\frac{1}{2}|\psi \rangle \langle \psi|$, for which the solution can be qualitatively described. For such a totally uncoherent state there are no oscillations and the intensity is the sum of the intensities produced with one of the slits closed. In the presence of matter the curve is rather broad and flat, because each target disperses the particle in all directions, thus increasing the odds that it hits the screen away from the center. Such a broad shape is only possible if the mixing piece itself is even broader, since  $\widetilde{\rho}^{coh}$ only reduces its size by a depletion factor $e^{-2Im(k')L}$.

This is already the end of the argument if one realizes that this $\rho^{mix}$  has to be the same either with the initial auxiliary state $\widetilde{\rho}(0)$ or with $|\phi \rangle+|\psi \rangle$, because  - and this is crucial - the crossed terms $|\phi \rangle \langle \psi|$  of the initial state do not intervene at all in the subsequent $\rho^{mix}$ evolution, as previously shown.                 

We thus see that the mixing part adds nothing but a constant background around the central region, that dies off very slowly as one moves away.

Summarizing, the observed interference pattern consists of a constant background, which entails the decoherence due to the medium, superposed to the oscillations of $\rho^{coh}$, also damped by the medium (see Fig. 2). The relative strength of the two contributions is dictated by unitarity, and one can roughly estimate the ratio of the oscillation amplitude over the background as:
\begin{eqnarray}
\frac{2 e^{-2Im(k')L}}{1-e^{-2Im(k')L}},
\end{eqnarray}
i.e.,  the smaller this number may get, the more invisible the interference fringes become.

\vspace{0.7cm} 

Finally, we would like to stress the observation that from the size of the tiny oscillations of the interference pattern in the central region, one can fit $e^{-2Im(k')}$. Precise fits in experiments of this kind thus provide a direct measurement of $Im(k')$, related to $Im(H)=-A^{\dagger}A$ in eq. (\ref{entenent lindblad}); i.e., the size of the Lindblad coefficients for the particle crossing a medium.

\begin{figure}[!ht]
\includegraphics[scale=0.55,angle=0]{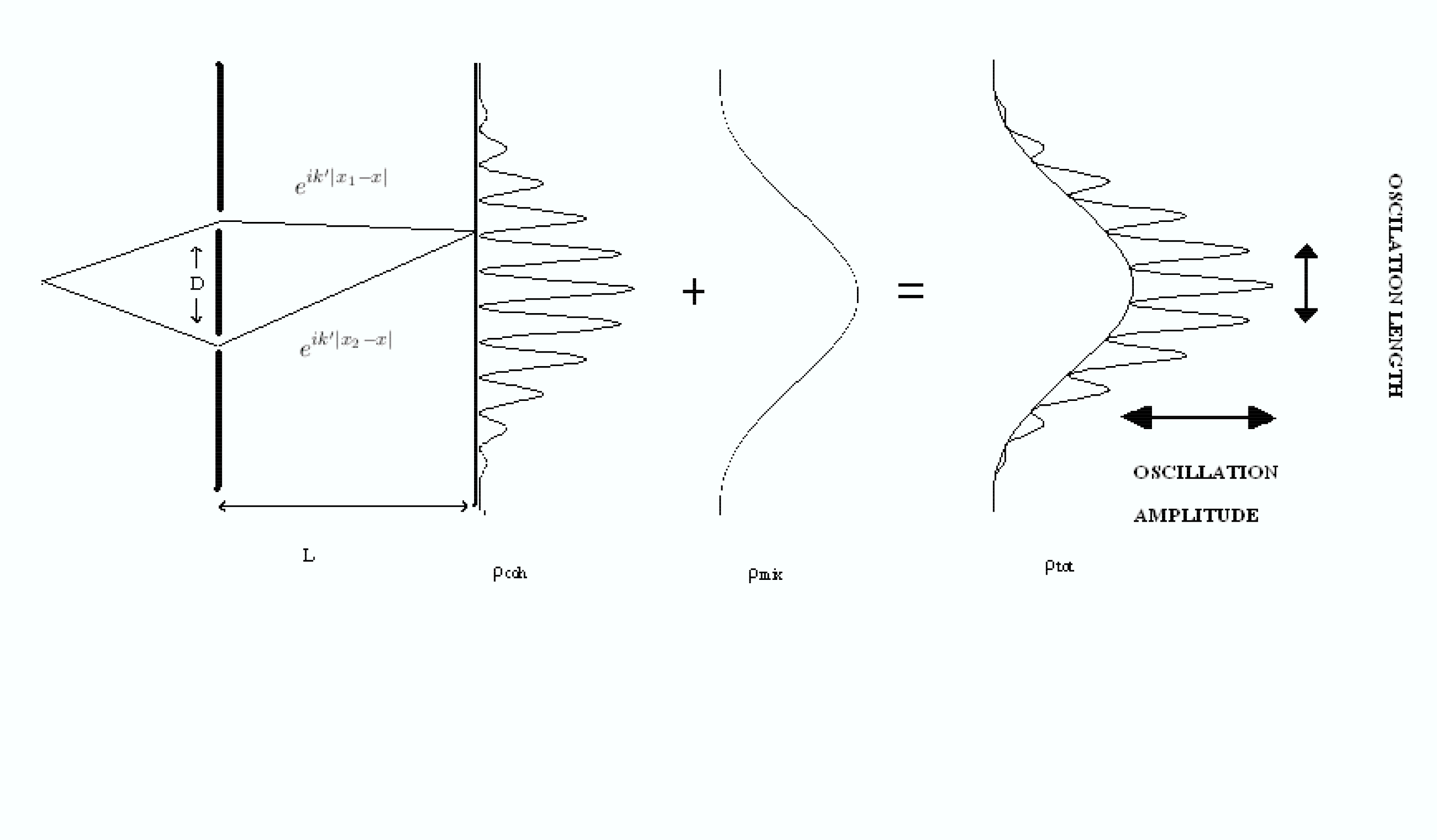}
\caption{The interference pattern wen a gas is present:}
\end{figure}

\section{Conclusions}

In this article we have addressed  a microscopic derivation of the Lindblad equation that describes the loss of quantum coherence of a particle that crosses a medium with which it interacts. This is done for a \textit{dilute} medium made out of targets that interact with a particle through a short range potential. In the case that the medium is a gas, we keep in the evolution equation terms up to second order in the potential. We also use the \textit{impulse approximation}, i.e., we approximate the scattering amplitude for the particle with the gas by the sum of the scattering amplitudes with each single target. We argue how the imaginary part of the refraction index given by Fermi is related to the Lindblad terms and is, therefore, due to decoherence. Finally, we show that when we apply our results to a double slit Young experiment, the characteristic interference patterns disappear due to the presence of the medium, a distinctive sign of decoherence.

We stress how the corrections due to the presence of matter split into different terms according to the two different mechanisms that produce decoherence: the \textit{footprint} mechanism, related to the entanglement with the medium; and the \textit{mixture} mechanism, related to the probabilistic knowledge of the medium state.

\section{Acknowledgments}

We thank A.Andrianov for helpful comments at an early stage of this work; E.Jan\'e for  discussions; and, specially, J.Taron i Roca for suggesting this subject to me, for his help and advise. I acknowledge finantial support from the following sources:  a grant by the \textit{Ministerio de Ciencia y Tecnolog'a} (Spain) MCYT FPA, 2001-3598; \textit{Generalitat de Catalunya }CIRIT, GC 2001SGR-00065; \textit{Euridice European Collaboration} HPRN-CT-2002-00311.


\begin{thebibliography}{99}

\bibitem{Fermi} E. Fermi, Nuclear Physics, The University of Chicago Press. (1951), pag. 201.

\bibitem{Jackson} J.D.Jackson, Clasical Electrodynamics,  Wiley (NY 1975), pag. 453.

\bibitem{MSW} S. P. Mikheyev,  A. Yu. Smirnov, Sov. J. Nucl. Phys. 42, 913 (1985).  

\bibitem{Wolfenstein} L. Wolfenstein, Phys. Rev. D 17, 1 (1978).

\bibitem{Bramon} B. Ancochea, A. Bramon, Phys. Lett. B347, 419 (1995);
A. Bramon, G. Garbarino, Phys Rev. Lett. 89, 160401 (2002).

\bibitem{kabir} P.K. Kabir, The CP Puzzle, Academic Press (1968).

\bibitem{Lindblad} G. Lindblad, Comm. Math. Phys. 48, 119 (1976).

\bibitem{Louisell} W. H. Louisell, Quantum Statistical Properties of Radiation, Wiley, New York (1973).

\bibitem{Cohen-Tannoudji}  C. Cohen-Tanoudji, J. Dupont-Roc, G. Grynberg, Atom-Photon Interactions, Wiley, New York (1992)

\bibitem{Stodolsky} R.A.Harris, L.Stodolsky, J. Chem. Phys. 74 2145
 (1981).

\bibitem{Zurek} W. H. Zurek, Report quant-ph/0105127; and references therein. 

\bibitem{Zeh} D. Giulini \textit{et al.}, Decoherence and the appearance of a classical world in Quantum Theory, Springer, Berlin (1996).   

\bibitem{peskin} P. Huet, M. Peskin, Nucl. Phys. B434 (1995) 3.


\bibitem{benatti} F. Benatti, R. Floreanini, Ann. Phys. 273 (1999) 58;
F. Benatti, R. Floreanini, Nucl. Phys. B488 (1997) 335; F. Benatti, R. Floreanini, Mod. Phys. Lett. A14 (1999) 1519.


\bibitem{taron} A. A. Andrianov, J. Taron, R. Tarrach,  Phys.Lett.B507:200-206,2001

\bibitem{review} S. M. Bilenky, Report hep-ph/0211462.   


\end{thebibliography}
\end{document}